\documentclass[12pt]{iopart}
\usepackage{iopams}
\usepackage{graphicx,psfrag}
\usepackage[hypertex]{hyperref}
\usepackage{txfonts}

\newcommand{\average}[1]{\left\langle{#1}\right\rangle}

\newcommand{\D}{\mathrm{d}}
\newcommand{\E}{\mathrm{e}}

\newcommand{\FF}{\mathcal{F}}

\newcommand{\LL}{\mathcal{L}}
\newcommand{\HH}{\mathcal{H}}

\newcommand{\tf}{t_{\mathrm{f}}}
\newcommand{\Nt}{{N_{\mathrm{ traj}}}}
\renewcommand{\br}{\mathbf{r}}
\newcommand{\tF}{t_{\mathrm{F}}}
\newcommand{\vF}{v_{\mathrm{F}}}
\newcommand{\zm}{z_{\mathrm{max}}}

\newcommand{\zeff}{v_{\mathrm{eff}}}
\newcommand{\kt}{k_\mathrm{B}T}
\newcommand{\F}{\mathrm{f}}

\newcommand{\p}[1]{\left({#1}\right)}
\newcommand{\pq}[1]{\left[{#1}\right]}
\newcommand{\pg}[1]{\left\{{#1}\right\}}
\renewcommand{\d}[2]{\frac{\partial #1}{\partial #2}}
\newcommand{\boldsym}[1]{\mbox{\boldmath${#1}$}}

\begin{document}
\title{%
Evaluation of free energy landscapes from
manipulation experiments}
\author{A. Imparato and L. Peliti}
\address{Dipartimento di Scienze Fisiche and INFN-Sezione di Napoli,
Universit\`a ``Federico II''\\
Complesso Universitario di Monte S. Angelo, I--80127 Napoli
(Italy)}
\ead{imparato@na.infn.it, peliti@na.infn.it}

\date{\today}

\begin{abstract} 
A fluctuation relation, which is an extended form of the Jarzynski equality, is introduced and discussed. We show how to apply this relation in order to evaluate the
 free energy landscape of simple systems. These systems are manipulated by varying the external field coupled with a systems' internal characteristic variable. Two different manipulation protocols are here considered: in the first case the external field is a linear function of time, in the second case it is a periodic function of  time. While for
simple mean field systems both the linear protocol and the oscillatory protocol 
provide a reliable estimate of the  free energy landscape, for a simple model of
homopolymer the oscillatory protocol turns out to be not reliable for this purpose. We then discuss the possibility of application of the method here presented to evaluate the free energy landscape of real systems, and the practical limitations that one can face in the realization of an experimental set-up. 

\end{abstract}
\noindent{\it Keywords}: Fluctuations (Theory)
, Energy landscapes (Theory),  Mechanical properties (DNA, RNA, membranes, bio-polymers) (Theory)  
\maketitle

\section{Introduction}
In the recent years, a number of manipulation experiments
have been performed with the aim of gathering information
on the equilibrium properties of complex molecular
systems, such as biopolymers~\cite{exp}. In particular,
in a class of these experiments the Jarzynski
equality (JE)~\cite{JE, Crooks1,Crooks,felix0} has been exploited in order
to evaluate the equilibrium free energy landscape
of the system, even if the system is no
more at equilibrium during the manipulation experiment.

In the present work, we wish to investigate under what
conditions the use of the JE in its several forms is effective
for the evaluation of the free energy landscape of a
small system. Indeed, the JE requires the evaluation of the average
of $\exp(-\beta W)$, where $\beta=1/\kt$ ($T$ is the temperature)
and $W$ is the work exerted on the system. This quantity
has a wide distribution, even if the distribution of $W$ is
comparatively narrow, and it is not clear \textit{a priori}
when  sufficient statistics for its evaluation can be mustered \cite{felix}.
Typically one wishes to evaluate the free energy landscape
of the system as a function of a collective coordinate $M$
which accessibly represents a semimacroscopic state
of the system. For example, in the case of pulling
experiments, one takes for $M$ the elongation of
the molecule. 
One thus needs to introduce extended forms of the JE in order
to evaluate the detailed free energy landscape as a function of
the internal characteristic variable.
By combining the JE and the 
 the histogram method
(cfr.~\cite{Ben, FeSw,HumSza,Seif1})  one is able to evaluate the
free energy of a constrained equilibrium state in which
the collective coordinate assumes a fixed value.

The article is organized as follows. In section \ref{uno}, we derive the 
extended form of the JE, which connects the work done on a manipulated system to its free energy landscape. In section \ref{histo_met} we review the histogram 
method and show how it can be exploited to evaluate free energy landscapes of
manipulated systems.
We then  apply the histogram method to evaluate the free energy landscape of a mean field Ising model, section \ref{ising}, and of a simple model of homopolymer, section \ref{model_h}. We discuss our results and conclude in section \ref{discuss}.
 
\section{The basic identity} \label{uno}
We shall now briefly recall the derivation of the basic identity
of the histogram method.

Let us consider a system described by the hamiltonian $\HH_0(x)$,
where $x$ identifies its microscopic state. Let us also assume
that the system is originally at equilibrium, so that its
distribution in phase space is described by the canonical
distribution function
\begin{equation}
  \rho_0(x)=\frac{\E^{-\beta \HH_0(x)}}{Z_0},
\end{equation}
where
\begin{equation}
  Z_0=\int \D x\,  \E^{-\beta \HH_0(x)}
\end{equation}
is the corresponding partition function. We shall assume that the
system is manipulated in the following way. Let $M(x)$ be an
observable quantity (a sufficiently smooth function of the
microscopic state of the system) and $U_\mu(M)$ a function of $M$,
dependent on a parameter $\mu$. In the initial state, without loss
of generality, we take $\mu=0$ and $U_{0}(M)\equiv 0$. The
manipulation protocol is defined by assigning a function $\mu(t)$,
($0\le t \le t_\F$), where $\mu(0)=0$ and $\mu(t_\F)=\mu$. The
time-dependent hamiltonian of the manipulated system is
$\HH_{\mu(t)}(x,t)=\HH_0(x)+U_{\mu(t)}(M(x))$. The work exerted on
the system up to time $t$ is a random quantity which depends on
the trajectory $x(t)$ that the system follows in phase space:
\begin{equation}
  W=\int_0^t \D t'\;\dot \mu(t')
    \left.\frac{\partial U_\mu(M(x(t'))}{\partial \mu}\right|_{\mu=\mu(t')}.
\end{equation}
The joint probability distribution $\Phi(x,W,t)$ of the microscopic
state $x$ and the accumulated work $W$ satisfies the
partial differential equation~\cite{HumSza,Seif,noi2,noi3}
\begin{equation}
  \frac{\partial \Phi}{\partial t}=\LL_{\mu(t)} \Phi-\dot \mu(t)
 \frac{\partial U_{\mu(t)}(M(x))}{\partial \mu}
 \frac{\partial \Phi}{\partial W}.
\end{equation}
Here, $\LL_\mu$ is an evolution operator whose equilibrium
distribution, for any $\mu$, is the canonical distribution defined
by the hamiltonian $\HH_{\mu}(x)=\HH_0(x)+U_{\mu}(M(x))$:
\begin{equation}
  \LL_\mu \,\frac{\E^{-\beta\HH_{\mu}(x)}}{Z_\mu}=0,
\end{equation}
where, of course, $Z_\mu=\int \D x\;\E^{-\beta \HH_{\mu}(x)}$. Let
us now define the generating function $\Psi(x,\lambda,t)$ of the
distribution of $W$ via the equation
\begin{equation}
  \Psi(x,\lambda,t)=\int \D W\;\E^{\lambda W}\,\Phi(x,W,t).
\end{equation}
Then $\Psi(x,\lambda,t)$ satisfies the differential equation
\begin{equation}
  \frac{\partial \Psi}{\partial t}=\LL_{\mu(t)} \Psi+\lambda \dot \mu(t)
  \frac{\partial U_{\mu(t)}(M(x))}{\partial \mu}\,\Psi.
\end{equation}
One can then easily check that, for $\lambda=-\beta$, the
corresponding equation and the initial condition
$\Psi(x,\lambda,t{=}0)=\rho_0(x)$ is identically satisfied by
\begin{equation} 
\Psi(x,-\beta,t)=\frac{\E^{-\beta \HH(x,t)}}{Z_0}.
\label{psi:eq}
\end{equation}
Integrating this relation on $x$ one obtains the usual form of
the JE:
\begin{equation}
 \average{\E^{-\beta W}}_t=\int \D x\int \D W\;\E^{-\beta W}\,
  \Phi(x,W,t)=\frac{Z_{\mu(t)}}{Z_0}
  =\exp\left[-\beta\left(F_{\mu(t)}-F_0\right)\right]
\label{je}
\end{equation}
Here $Z_\mu$ is the partition function corresponding to the
hamiltonian $\HH_0(x)+U_\mu(M(x))$, and $F_\mu=-\kt \ln Z_\mu$
is the corresponding free energy.
A more general relation is obtained if we multiply both
sides of eq.~(\ref{psi:eq}) by $\delta(M-M(x))$ before integrating:
\begin{equation}
\average{\delta(M-M(x))\E^{-\beta W}}_t=
  \int \D x\;\delta(M-M(x))\frac{\E^{-\beta \HH(x,t)}}{Z_0}
  =\E^{-\beta \pq{\FF_0(M)+U_{\mu(t)}(M)-F_0}}.
\label{constrained:eq}
\end{equation}
Here $\FF_0(M)$ is the free energy of a constrained ensemble, in
which the value $M(x)$ is fixed at $M$:
\begin{equation}
  \FF_0(M)=-\kt \ln\int \D x\;\delta(M-M(x))\,\E^{-\beta \HH_0(x)}.
\end{equation}
Note that eq.~(\ref{constrained:eq}) corresponds to eq.~(21) in Crooks
 (ref.~\cite{Crooks}), with the substitution $f(x)=\delta (M-M(x))$. It generalizes the expression 
 by Hummer and Szabo (eq. 4, ref.~\cite{HumSza}) to the case in which the state of the system is represented by a collective coordinate.
In this case the expression on the rhs of eq.~(\ref{constrained:eq}) involves
a free energy function rather than a microscopic hamiltonian.

By multiplying both sides of eq.~(\ref{constrained:eq}) by
$\E^{\beta U_{\mu(t)}(M)}$, we obtain the basic identity of the
histogram method:
\begin{equation}
  \E^{\beta U_{\mu(t)}(M)}\average{\delta(M-M(x))\E^{-\beta W}}_t
  =\E^{-\beta \pq{\FF_0(M)-F_0}}.
\label{sample}
\end{equation}

Equation~(\ref{sample}) provides thus a method to evaluate the
unperturbed free energy $\FF_0(M)$ as long as one has a reliable
estimate of the lhs of this equation. Note that that the quantity
on the rhs of eq.~(\ref{sample}) is a time independent quantity,
and thus an improved estimate of $\FF_0(M)$ can be obtained by
sampling the rhs of eq.~(\ref{sample}) at different time $t$ along
the manipulation process. The problem is that the quantities so
obtained are not equally distributed, and so, their statistical
treatment has to be performed conveniently, as described in the
next section.

Equation (\ref{sample}) can be viewed as an extension of the JE
(\ref{je}). This last equation provides a method to evaluate the
equilibrium free energy difference $\Delta F_t$ between the two
thermodynamical states characterized by the external parameter
values $\mu(t)$ and $\mu(0)$: one can in fact evaluate the
quantity $\Delta F^*_t$ defined by the following equation
\begin{equation}
    \E^{-\beta\Delta  F^*_t}=
    \frac{1}{\Nt} \sum^\Nt_{i=1} \E^{-\beta W^i_t}\equiv \overline{\E^{-\beta W_t}}.
\label{fstar}
\end{equation}
The best estimate for $\Delta F_t$ will thus be given by $\Delta
F_t\simeq \Delta F^*_t$.

\section{Histogram method for the evaluation of the free energy landscape} \label{histo_met}
Let us assume that we have $n$ random variables $x_i$, $i=1,\dots,
n$, which are not identically distributed, but have the same
average value $\average {x_i}=X$. We wish to estimate $X$ from a
given sample $\pg{x_i}$ of the $x_i$'s. Let us write the quantity
$x_i$ as a product of a random variable $\xi_i$ and of a
non-fluctuating factor $a_i$,
\begin{equation}
    x_i=\xi_i a_i.
\end{equation}
One can obtain an estimate $X_p$ of $X$ from the set of data
$\pg{x_i}$ by a linear combination
\begin{equation}
    X_p=\sum_{i=1}^n p_i x_i=\sum_{i=1}^n p_i \xi_i a_i,
\end{equation}
where the coefficients $p_i$ satisfy
\begin{equation}
    p_i\ge 0; \qquad \sum_{i=1}^n p_i=1.
\label{constr}
\end{equation}
The best estimate of $X$ is obtaining by minimizing the variance
\begin{equation}
    \Delta X^2_p=\average{X_p^2}-\average{X_p}^2
\end{equation}
of the fluctuating quantity $X_p$, under the constraints
(\ref{constr}). If one has
\begin{equation}
    \sigma^2_i=\average {\xi_i^2} -\average {\xi_i}^2,
\end{equation}
the variance of $X_p$ is given by
\begin{equation}
    \Delta X_p^2 =\sum_{i=1}^N p_i^2 a_i^2 \sigma_i^2.
\end{equation}
By minimizing $\Delta X_p^2$, we thus obtain the following
expression for the coefficients $p_i$:
\begin{equation}
    p_i=\frac{\lambda}{a_i^2 \sigma_i^2},
\end{equation}
where $\lambda$ is a Lagrange multiplier, which is fixed by the
normalization condition of the coefficients $p_i$:
\begin{equation}
    \lambda^{-1}=\sum_{i=1}^n \frac{1}{a_i^2 \sigma_i^2}.
\label{defl}
\end{equation}
The best estimate of $X$ is thus given by
\begin{equation}
    X^*_p=\lambda \sum_{i=1}^n \frac{\xi_i}{a_i \sigma_i^2},
\label{xsp}
\end{equation}
where $\lambda$ is given by eq.~(\ref{defl}).

As discussed above, we want to evaluate the rhs of
equation~(\ref{sample}) by sampling the experimentally accessible
quantity which appears on the lhs of the same equation. We thus
consider a number $N_\mathrm{traj}$ of repetitions of the
experiment, and sample the corresponding trajectories at discrete
times $t_j=j\,\delta t$. We also divide the interval of possible
values of $M$ into bins $B_\ell=[M_\ell,M_\ell+\delta M_\ell)$.
Let us define the random variable
\begin{eqnarray}
r(M_\ell,t_j)&=&Z_0\, \E^{\beta U_{\mu(t_j)}(M_\ell)}
\overline{\theta_\ell(M(t_j)) \E^{-\beta W}}\nonumber\\
&=& Z_0\, \E^{\beta U_{\mu(t)}(M_\ell)}\; \frac{1}{\Nt}\sum^\Nt_{k=1}\theta_\ell(M^k_{t_j})
\,\E^{-\beta W^k_{t_j}},
\end{eqnarray}
where the sum runs over $\Nt$ independent repetitions of the
manipulation process and $M^k_{t_j}$ is the value of the variable
$M$ along the $k$-th trajectory, at sampling time $t_j$. We have
introduced the characteristic function $\theta_\ell(M)$ of the
$\ell$-th bin:
\begin{equation}\label{characteristic:eq}
    \theta_\ell(M)=\left\{%
\begin{array}{ll}
    1, & \hbox{if $M_\ell\le M<M_\ell+\delta M_\ell$ ;} \\
    0, & \hbox{otherwise.} \\
\end{array}%
\right.
\end{equation}
Let
\begin{equation}
\rho(M_\ell,t_j)=\frac{1}{\Nt}\sum^\Nt_{k=1}\theta_\ell(M^k_{t_j})\,
\E^{-\beta W^k_{t_j}}\label{defrho}
\end{equation}
define the stochastic part of the variable $r(M_\ell,t_j)$. We use
$r(M_\ell,t_j)$ to estimate the quantity
\begin{equation}\label{boltzfac:eq}
    \Delta R(M_\ell) =\exp\pq{-\beta \FF_0(M_\ell)}\,\delta
    M_\ell.
\end{equation}
According to eq.~(\ref{xsp}) the best estimate for $\Delta
R(M_\ell)$ is given by
\begin{equation}
\Delta R^*(M_\ell)=\lambda\sum_j
\frac{\rho(M_\ell,t_j)}{a_{t_j}(M_\ell) \sigma^2_{t_j}(M_\ell)},
\label{defRs}
\end{equation}
where $a_t(M_\ell)$ is given by
\begin{equation}
a_{t_j}(M_\ell)=\E^{\beta U_{\mu(t_j)}(M_\ell)},
\end{equation}
$\sigma^2_{t_j}(M_\ell)$ is given by
\begin{eqnarray}
 \sigma^2_{t_j}(M_\ell)&=&\average{\rho^2(M_\ell,t_j)}-\average{\rho(M_\ell,t_j)}^2\nonumber\\
 &=& \frac{1}{\Nt^2}\sum^\Nt_{k=1}  \average{\theta_\ell(M^k_{t_j}) \E^{-2\beta W^k_{t_j}}}-\average{\frac{1}{\Nt}\sum^\Nt_{k=1}\theta_\ell(M^k_{t_j}) \E^{-\beta W^k_{t_j}}}^2 ,
\end{eqnarray}
and $\lambda$ is defined by the normalization condition
(\ref{defl}). In ref.~\cite{Seif1}, $\sigma^2_{t}(M)$ is taken to
be
\begin{equation}
    \sigma^2_{t}(M)=\frac{\overline{\E^{-\beta W_t}}}{a_t(M)}.
\label{sigSe}
\end{equation}
Note that the quantity appearing on the numerator of
eq.~(\ref{sigSe}) does not necessarily satisfy the JE, since the
mean is taken over a finite number of trajectories.

Note also that the rhs of eq.~(\ref{fstar}) is equal to the numerator
of the fraction appearing  on the rhs of eq.~(\ref{sigSe}).

\section{Evaluation of the free energy landscape of a mean field system}\label{ising}
In the following we apply the histogram method, discussed in the previous section,  to probe the free energy landscape of
a known system, namely an Ising model in  mean-field approximation, whose
unperturbed free energy reads
\begin{equation}
\FF_0(M)=-\frac {J}{2 N} M^2 - T S(M),
\end{equation}
where where $S(M)$ is the usual entropy for an Ising paramagnet,
\begin{equation}
  S(M)=-k_\mathrm{B}\left[\left(\frac{N+M}{2}\right)
    \log\left(\frac{N+M}{2}\right)+\left(\frac{N-M}{2}\right)
    \log\left(\frac{N-M}{2}\right)\right].
  \label{entropy:eq}
\end{equation}
By using such a mean-field model, we can test how the different system's parameters affect the effectiveness of the histogram method to evaluate the free energy landscape. In particular we  analyze the effect of changing the system size, the interaction parameter $J_0$, the manipulation protocol, and the manipulation rate.
The method described here can be easily generalized to  systems characterized 
by any given free energy function, at least as long as the space of collective variables remains of small dimensionality.

The free energy landscape will be probed by applying an external magnetic field $h$, which is manipulated according to a given protocol $h(t)$.
We assume that the system evolves according to  Langevin dynamics, and
thus the manipulation process can be simulated by numerically integrating the Langevin equation
\begin{equation}
\d{M}{t}=-\beta \nu_0\d{\FF(M)}{M}+\eta(t),
\label{lan:eq}
\end{equation}
with
\begin{equation}
\average{\eta(t)\eta(t')}=2 \nu_0\delta(t-t'),
\end{equation}
and where the total free energy $\FF(M)$ is given by
\begin{equation}
\FF(M)=\FF_0(M)-hM
\end{equation}
The Langevin equation (\ref{lan:eq}) can be integrated
using the Heun algorithm \cite{Greiner}.

As discussed in the previous section, by sampling the quantity
$\rho(M,t)$, eq.~(\ref{defrho}), we obtain the best estimate for
$\FF_0(M)=-k_B T\ln R^*(M)$, where $R^*(M)$ is defined by
eq.~(\ref{defRs}). Here we adopt the expression (\ref{sigSe}) for
$\sigma^2_t(M)$. Let us define the magnetization per spin $m=M/N$,
$N$ being the system size, then the quantity $f^*_0(m)$ defined as
\begin{equation}
f^*_0(m)=-\frac{k_B T}{N}\ln R^*(M),
\label{f0star}
\end{equation}
will indicate in the following the estimated free energy per spin,
for a given manipulation  protocol. In order to quantify the
quality of each estimate we divide the interval of values of $m$
$\pq{-1,1}$ into $N_m$ bins, and define the {\it distance}
function
\begin{equation}
d\equiv\frac {1}{N_m} \sum_{i=0}^{N_m} \pq{f_0(m_i)-f_0 ^*(m_i)}^2.
\label{defd}
\end{equation}
As we will see in the next subsections,  the quality of the
estimate of $\FF_0(M)$ via eq.~(\ref{f0star}) is strictly
connected to the the quality of the estimate of $\Delta F_t$ via
eq. (\ref{fstar}).

In the following, we fix the energy scale and the time scale by
taking $\beta=1$ and $\nu_0=1$.

\subsection{Linear protocol}\label{linp}
We first consider a linear protocol, which reads
\begin{equation}
h(t)=h_0+\frac{h_1-h_0}{\tf} t.
\label{hlin}
\end{equation}
\begin{figure}[h]
\center
\psfrag{t2}[cl][cl][.8]{$\tf=2$}
\psfrag{t10}[cl][cl][.8]{$\tf=10$}
\psfrag{m}[ct][ct][1.]{$m$}
\includegraphics[width=8cm]{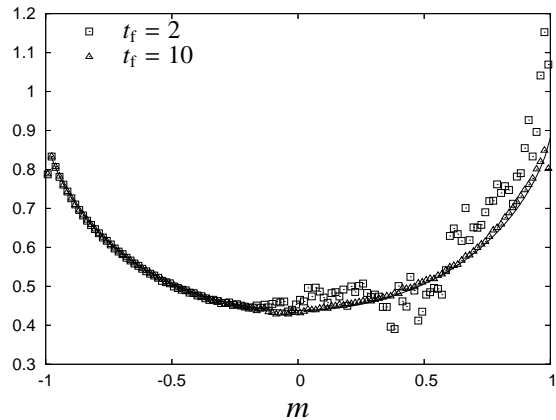}
\caption{Comparison of the free energy landscape
$f_0(m)=\FF_0(M)/N$ (full line) with $J_0=0.5$, with the results
of the simulations described in the text.  The external magnetic
field is varied according to the protocol (\ref{hlin}) with
$h_1=-h_0=1$, and $\tf=2,10$. The system size is $N=10$ and for
each value of $\tf$, $\Nt=10^4$ samples of the process are taken.}
\label{confN10}
\end{figure}

\begin{figure}[h]
\center
\psfrag{t2}[cl][cl][.8]{$\tf=2$}
\psfrag{t10}[cl][cl][.8]{$\tf=10$}
\psfrag{ht}[ct][ct][1.]{$h(t)$}
\includegraphics[width=8cm]{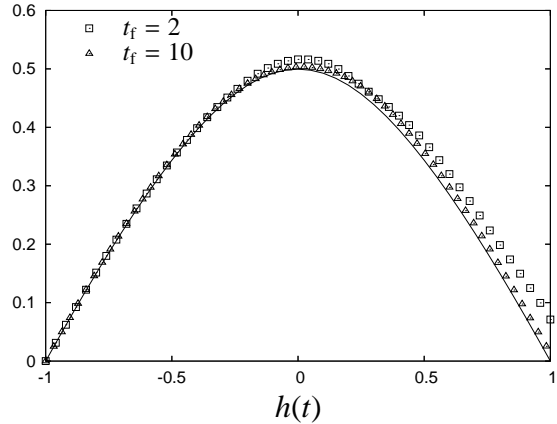}
\caption{Comparison of the equilibrium free energy difference
$\Delta f_t=f_{h(t)}-f_{h_0}$ as a function of $h(t)$ (full line),
with the quantity $F^*_t/N$, as defined by eq. (\ref{fstar}),
obtained with the simulations described in the text. In these
simulations  the external magnetic field is varied according to
the linear protocol (\ref{hlin}) with $h_1=-h_0=1$, and
$\tf=2,10$. The mean field interaction parameter $J_0$ is taken to
be $J_0=0.5$. The system size is $N=10$ and for each value of
$\tf$, $\Nt=10^4$ samples of the process are taken.}
\label{confJN10}
\end{figure}
\begin{figure}[h]
\center
\psfrag{t2}[cl][cl][.8]{$\tf=2$}
\psfrag{t10}[cl][cl][.8]{$\tf=10$}
\psfrag{t30}[cl][cl][.8]{$\tf=30$}
\psfrag{m}[ct][ct][1.]{$m$}
\includegraphics[width=8cm]{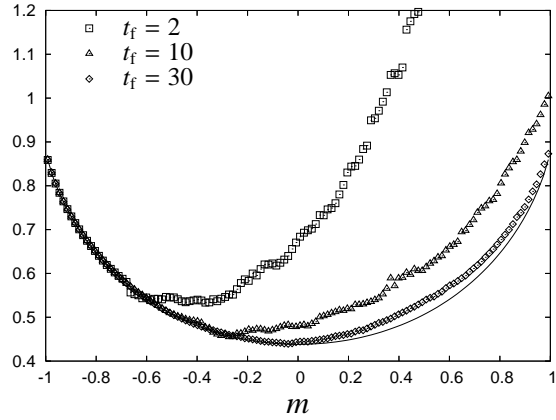}
\caption{Comparison of the free energy landscape
$f_0(m)=\FF_0(M)/N$ (full line)with $J_0=0.5$, with the results of
the simulations described in the text.  The external magnetic
field is varied according to the protocol (\ref{hlin}) with
$h_1=-h_0=1$, and $\tf=2,10,30$. The system size is $N=100$ and for
each value of $\tf$, $\Nt=10^4$ samples of the process are taken.}
\label{confN100}
\end{figure}
\begin{figure}[h]
\center
\psfrag{t2}[cl][cl][.8]{$\tf=2$}
\psfrag{t10}[cl][cl][.8]{$\tf=10$}
\psfrag{t30}[cl][cl][.8]{$\tf=30$}
\psfrag{ht}[ct][ct][1.]{$h(t)$}
\includegraphics[width=8cm]{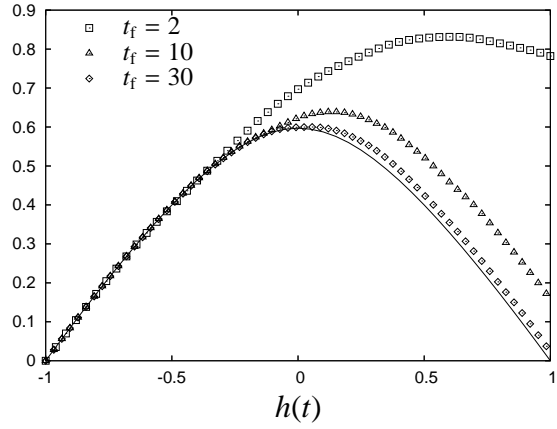}
\caption{Comparison of the equilibrium free energy difference
$\Delta f_t=f_{h(t)}-f_{h_0}$ as a function of $h(t)$ (full line),
with the quantity $F^*_t/N$, as defined by eq. (\ref{fstar}),
obtained with the simulations described in the text. The mean
field interaction parameter $J_0$ is taken to be $J_0=0.5$. In
these simulations  the external magnetic field is varied according
to the linear protocol (\ref{hlin}) with $h_1=-h_0=1$, and
$\tf=2,10,30$. The system size is $N=100$ and for each value of
$\tf$, $\Nt=10^4$ samples of the process are taken.}
\label{confJN100}
\end{figure}

In figure \ref{confN10}, the expected free energy per spin
$f_0(m)=\FF_0(M)/N$, is plotted together with the estimated free
energy $f^*_0(m)$, obtained with the method described above, for
two values of the final time (inverse manipulation rate) $\tf$.
This figure clearly shows that only the slowest process, with
$\tf=10$, gives a correct shape of the free energy  $f_0(m)$ for
any value of $m$. 
In figure
\ref{confJN10}, the quantity $\Delta F^*_t/N$, is plotted for the
two manipulation rates, together with the equilibrium free energy
difference $\Delta f_t=f_{h(t)}-f_{h_0}$, as functions of the
external magnetic field $h(t)$. Comparison of fig.~\ref{confN10}
with fig.~\ref{confJN10} gives strong evidence that the
effectiveness of the method here discussed for the reconstruction
of the free energy landscape is strictly related to its
effectiveness in evaluating the equilibrium free energy difference
by using the JE. If the manipulation protocol is such that  the
free energy landscape is successfully reconstructed, then the
estimate of the free energy difference, as given by
eq.~(\ref{fstar}) is close to its expected value $\Delta F_t$. On
the other hand, one cannot expect $\FF_0(M)$ to be reliably
evaluated if the total free energy difference $\Delta F_t$ is
poorly estimated. This conclusion is confirmed by changing the
system parameters, e.g., by increasing the system size $N$. In
figures \ref{confN100} and \ref{confJN100} we plot the same
quantities, namely $f_0^*(m)$ and $\Delta F^*_t/N$ for a larger
system, with $N=100$ spins. Also in this case the JE is effective
in giving an accurate estimate of the free energy difference
$\Delta F_t$, if the protocol is slow enough for the free energy
landscape to be reconstructed correctly.

\subsection{Oscillatory protocol}
In this subsection, the mean field system is manipulated according
to the oscillatory protocol
\begin{equation}
h(t)=h_0\sin (2 \pi \nu t),\qquad 0\le t\le\tf. \label{hosc}
\end{equation}
\begin{figure}[h]
\center
\psfrag{d}[ct][ct][1.]{$d$}
\psfrag{o}[ct][ct][1.]{$\omega$}
\psfrag{nu}[ct][ct][1.]{$\nu$}
\psfrag{0.5}[cl][cl][.8]{$J_0=0.5$}
\psfrag{1.1}[cl][cl][.8]{$J_0=1.1$}
\includegraphics[width=8cm]{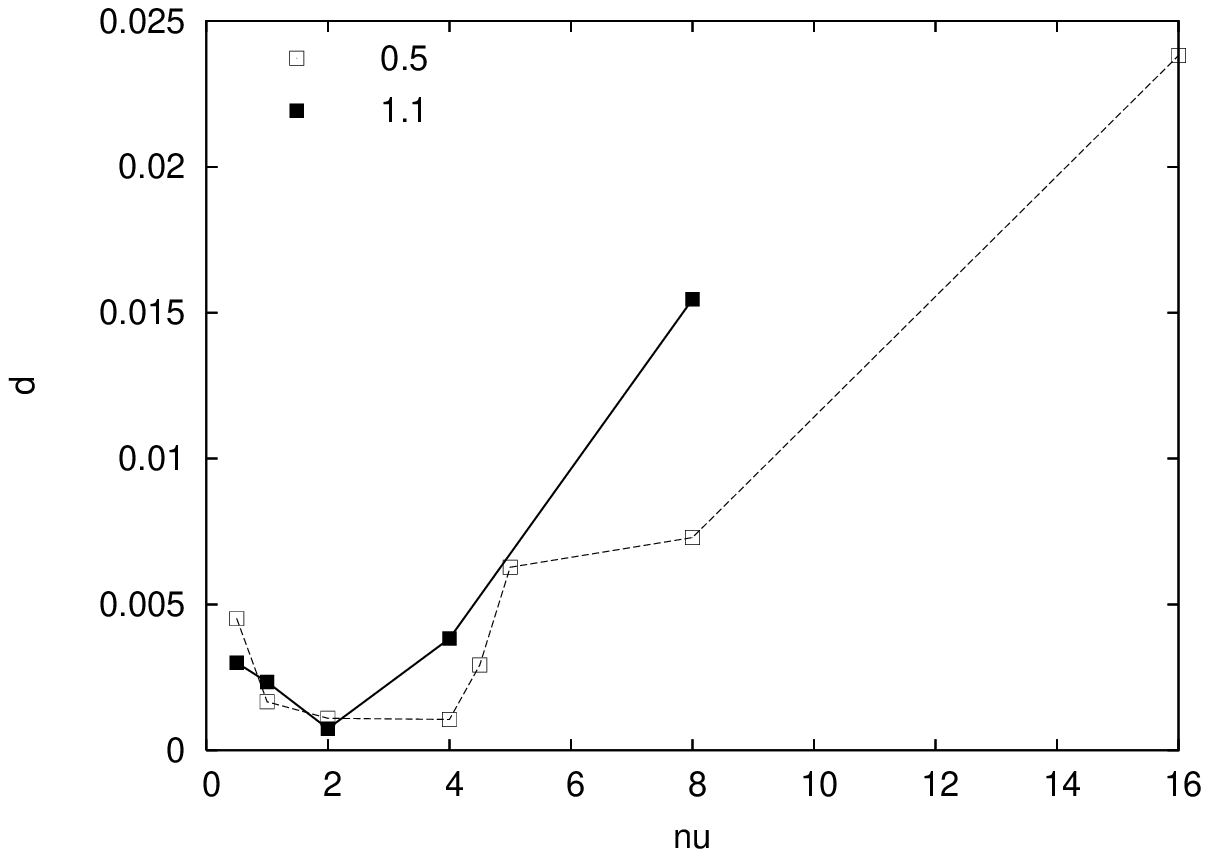}
\caption{Distance function $d$, as defined by eq.~(\ref{defd}), as
a function of the frequency $\nu$, for the oscillatory
manipulation protocol (\ref{hosc}). The size of the system is
$N=10$, and the number of trajectories is $\Nt=10^4$. The line is
a guide to the eye.} \label{d_omega_tf2}
\end{figure}
We set $\tf=2$, and take two values ($J_0=0.5,\, 1.1$) of the
interaction parameter $J_0$. The frequency $\nu$ is varied from
$1/2$ up to $16$. The function $d$, as defined by eq.~(\ref{defd})
is plotted in figure \ref{d_omega_tf2} as a function of $\nu$, for
the two values of $J_0$ here considered.

\begin{figure}[h]
\center
\psfrag{3.14}[cr][cr][.8]{$\nu= 1/2$}
\psfrag{12.56}[cr][cr][.8]{$\nu=2$}
\psfrag{25.13}[cr][cr][.8]{$\nu=4$}
\psfrag{50.26}[cr][cr][.8]{$\nu=8$}
\psfrag{100.531}[cr][cr][.8]{$\nu=16$}
\psfrag{m}[ct][ct][1.]{$m$}
\includegraphics[width=8cm]{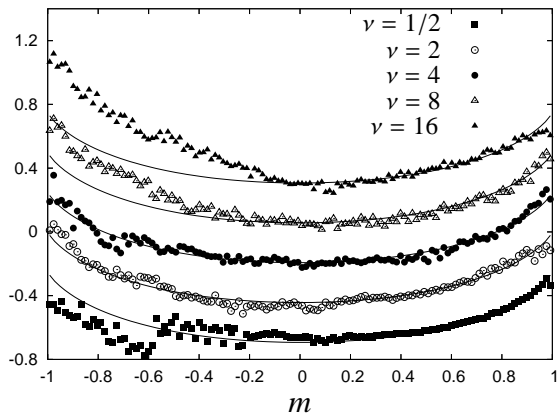}
\caption{Plot of the estimated free energy landscape $f_0^*(m)$,
obtained with the oscillatory protocol (\ref{hosc}), for different
values of the frequency $\nu$, and with $J_0=0.5$. The sets of
data are shifted to improve the clarity of the plot. The full
lines correspond to the expected free energy landscape $f_0(m)$.}
\label{confronto_omega}
\end{figure}

\begin{figure}[h]
\center
\psfrag{a}[ct][ct][1.]{(a)}
\psfrag{b}[ct][ct][1.]{(b)}
\psfrag{3.14}[cr][cr][.8]{$\nu= 1/2$}
\psfrag{12.56}[cr][cr][.8]{$\nu=2$}
\psfrag{25.13}[cr][cr][.8]{$\nu=4$}
\psfrag{50.26}[cr][cr][.8]{$\nu=8$}
\psfrag{h}[ct][ct][1.]{$h(t)$}
\includegraphics[width=8cm]{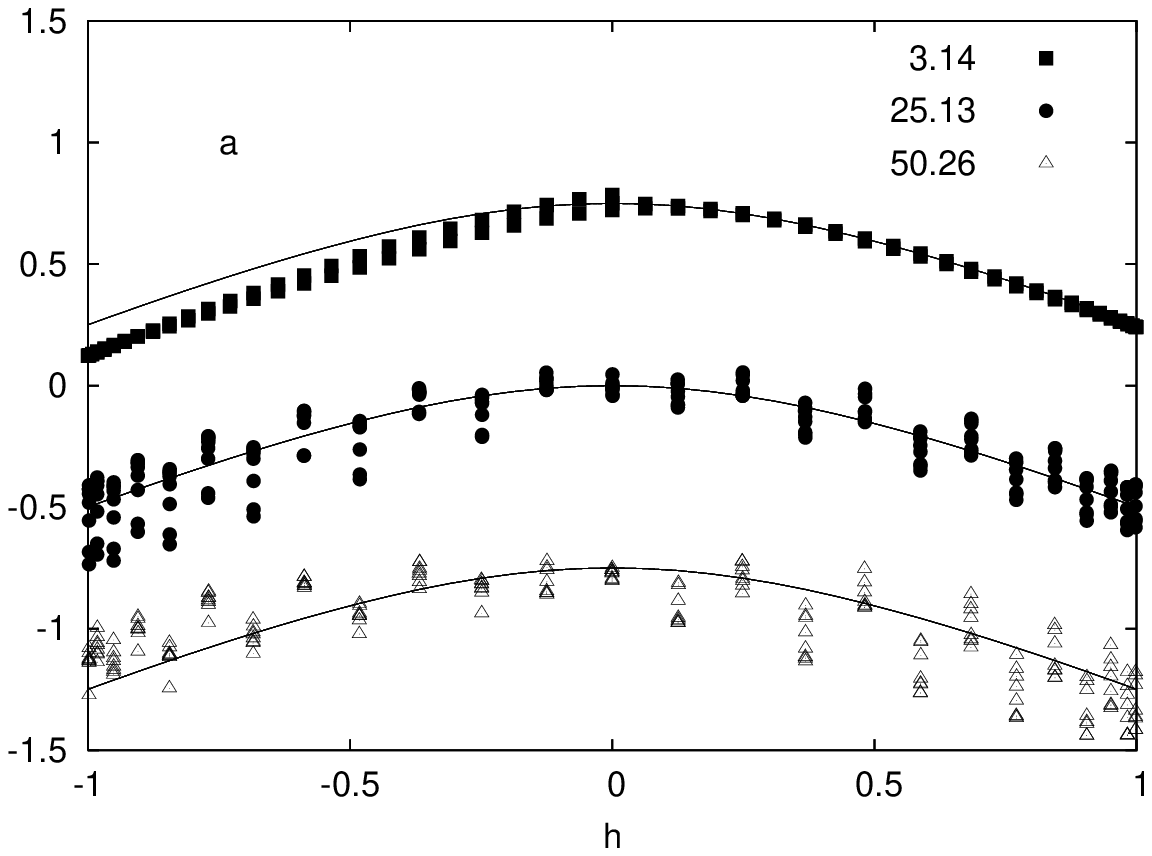}
\psfrag{h}[ct][ct][1.]{$h$}
\includegraphics[width=8cm]{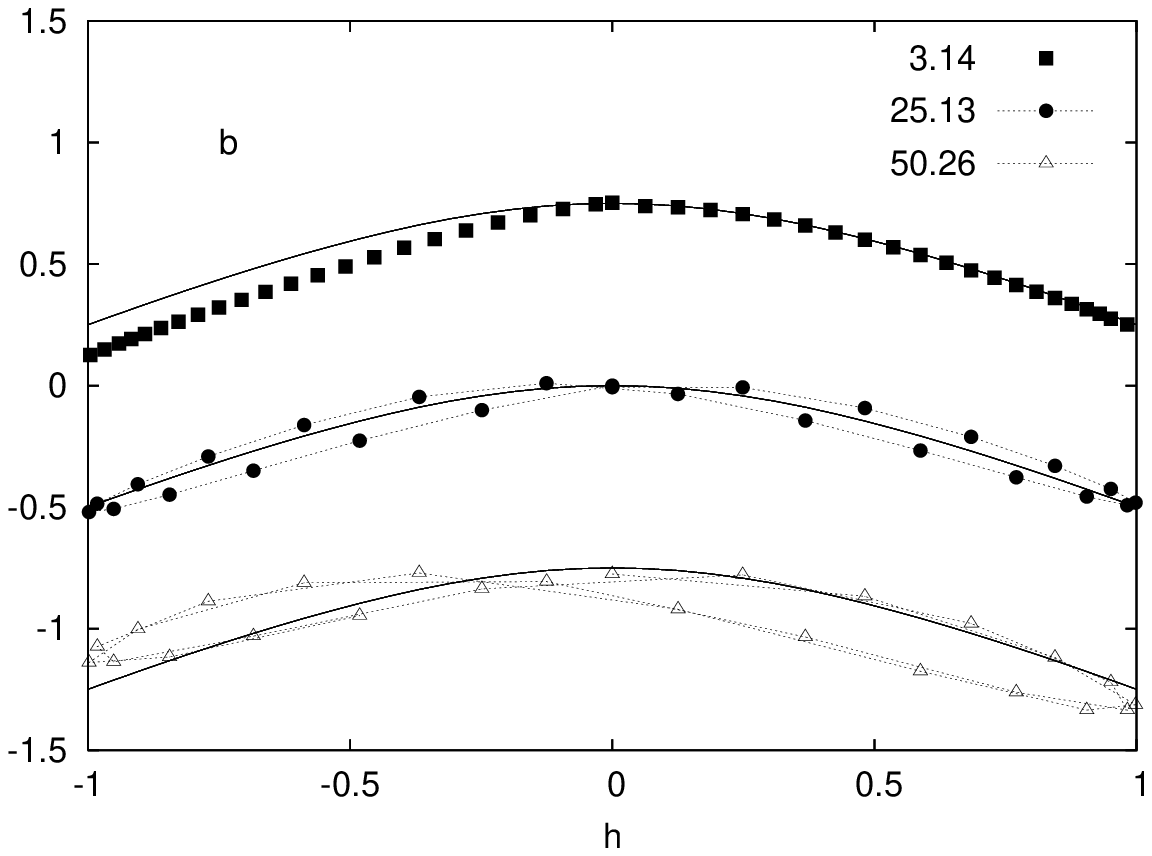}
\caption{Comparison of the equilibrium free energy difference
$\Delta f_t=f_{h(t)}-f_{h_0}$ as a function of $h(t)$ (full line),
with the quantity $F^*_t/N$, as defined by eq. (\ref{fstar}),
obtained with the simulations described in the text. In these
simulations  the external magnetic field is varied according to
the oscillatory protocol (\ref{hosc}) with $h_0=1$, and $\tf=2$.
The mean field interaction parameter $J_0$ is taken to be
$J_0=0.5$. The system size is $N=10$ and  $\Nt=10^4$ samples of
the process are taken. The sets of data are shifted to improve the
clarity of the plot. Panel (b): mean value of of $F^*/N$, obtained
by averaging the contributions to this quantity for a given value
of $h$, as plotted in panel (a). The dotted lines are guides to
the eye.} \label{confrontoj_omega}
\end{figure}

In figure \ref{confronto_omega} the expected free energy per spin
$f_0(m)$ is compared to the evaluated free energy for some of the
values of $\nu$ here considered, for the $J_0=0.5$ case.
Comparison of figures  \ref{d_omega_tf2} and \ref{confronto_omega}
clearly indicates that the optimal frequency for the
reconstruction of the free energy landscape, with $J_0=0.5$, does
not correspond to the smallest one but rather to $\nu\simeq 4 $.
The results discussed in subsection \ref{linp} suggest that the
estimate of the free energy landscape is optimal for a
manipulation protocol such that the estimate of the free energy
difference $\Delta f$ given by the JE is optimal. We thus consider
the estimated free energy difference $\Delta F^*_t$, as defined by
eq.(\ref{fstar}), for different values of the manipulation
protocol frequency $\nu$, and compare it to its expected value,
see  fig.~\ref{confrontoj_omega}(a). Since we are considering an
oscillating protocol here, eq.~(\ref{hosc}), for a given value of
$h$, there will be several estimates of $\Delta f_t$ for different
times $t$ separated by the protocol period $1/\nu$ .
In figure \ref{confrontoj_omega}(b), the mean value of $F^*_t/N$,
obtained by averaging over these different contributions for a
given value of $h$ is plotted. As for the reconstruction of the
energy landscape $f^*_0(m)$, the results shown in this last figure
indicate that the optimal frequency value, for estimating the free
energy difference $\Delta F_t$ is $\nu\simeq 4$.
\begin{figure}[h]
\center
\psfrag{3.14}[cr][cr][.8]{$\nu= 1/2$}
\psfrag{12.56}[cr][cr][.8]{$\nu=2$}
\psfrag{25.13}[cr][cr][.8]{$\nu=4$}
\psfrag{50.26}[cr][cr][.8]{$\nu=8$}
\psfrag{100.531}[cr][cr][.8]{$\nu=16$}
\psfrag{h}[ct][ct][1.]{$m$}
\psfrag{m}[ct][ct][1.]{$m$}
\includegraphics[width=8cm]{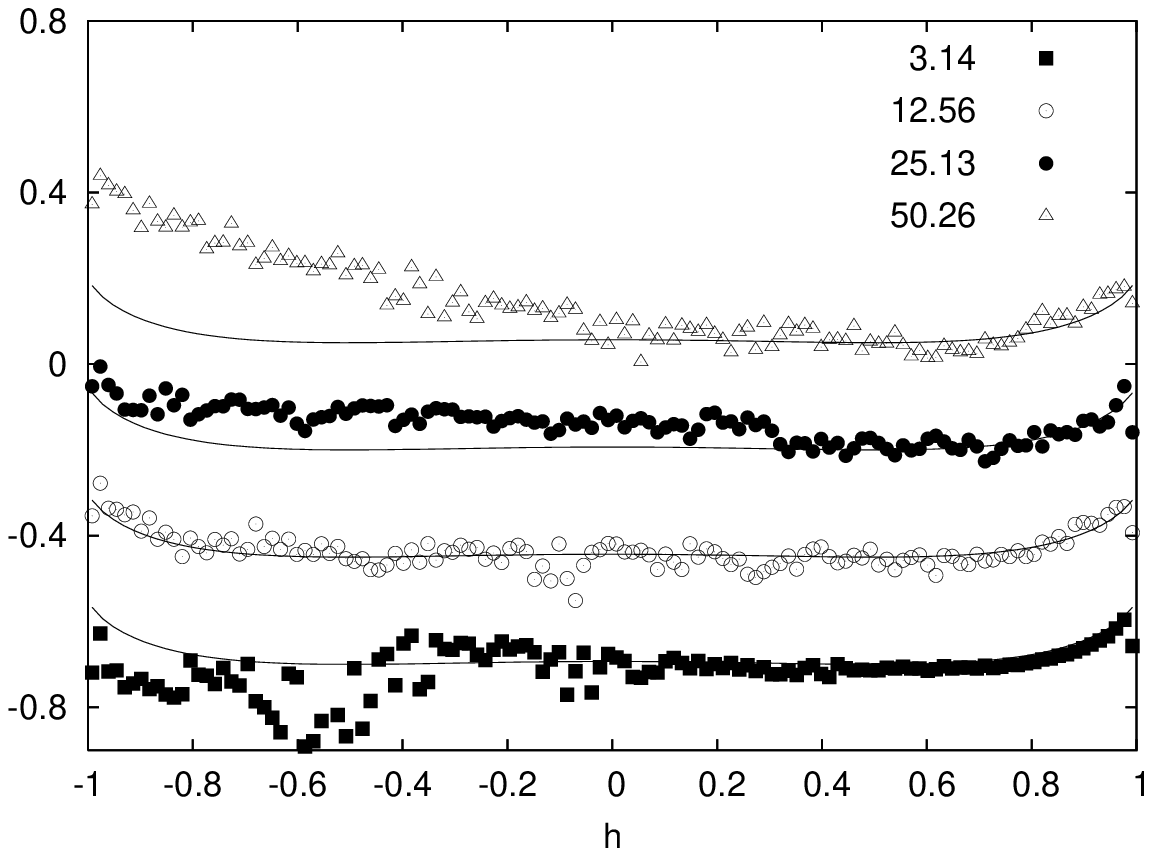}
\caption{Plot of the estimated free energy landscape $f_0^*(m)$,
obtained with the oscillatory protocol (\ref{hosc}), for different
values of the frequency $\nu$ and with $J_0=1.1$. The sets of data
are shifted to improve the clarity of the plot. The full lines
correspond to the expected free energy landscape $f_0(m)$.}
\label{confronto_omega_J1.1}
\end{figure}

\begin{figure}[h]
\center
\psfrag{3.14}[cr][cr][.8]{$\nu= 1/2$}
\psfrag{12.56}[cr][cr][.8]{$\nu=2$}
\psfrag{25.13}[cr][cr][.8]{$\nu=4$}
\psfrag{50.26}[cr][cr][.8]{$\nu=8$}
\psfrag{a}[ct][ct][1.]{(a)}
\psfrag{b}[ct][ct][1.]{(b)}
\psfrag{h}[ct][ct][1.]{$h(t)$}
\includegraphics[width=8cm]{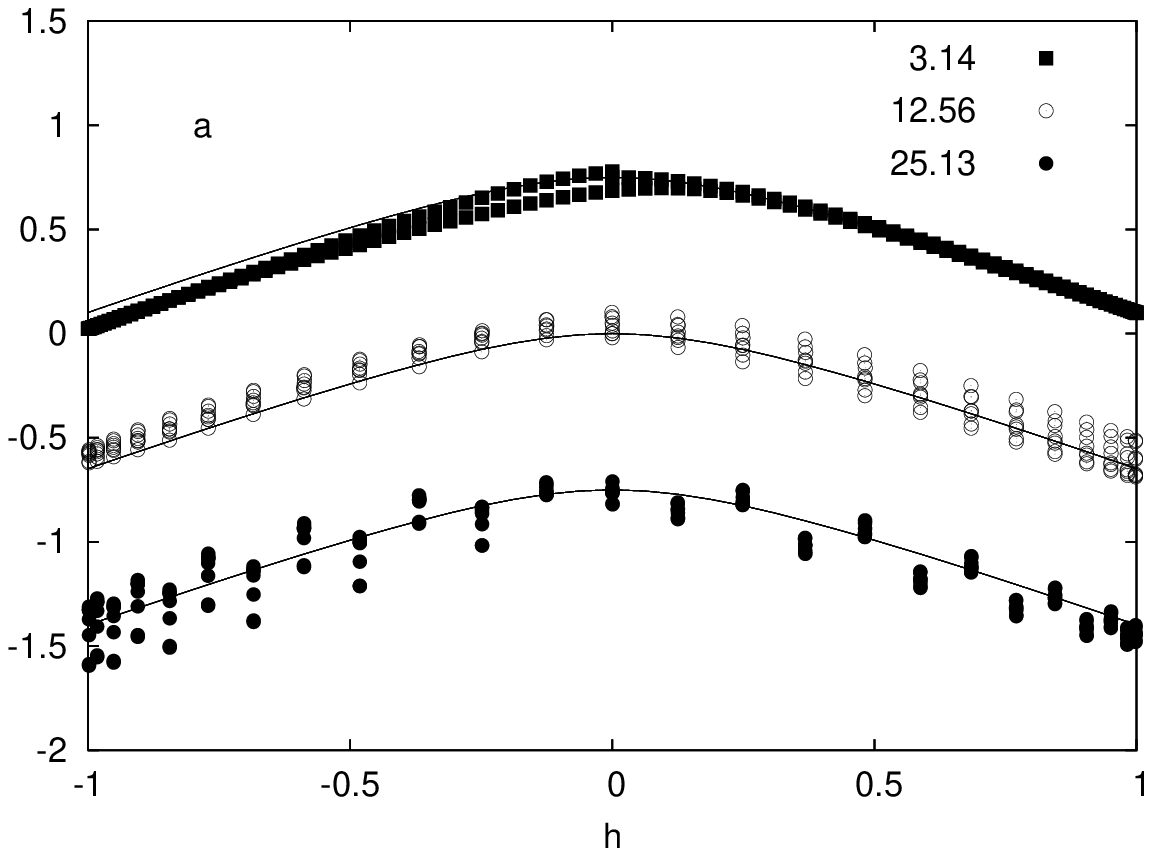}
\psfrag{h}[ct][ct][1.]{$h$}
\includegraphics[width=8cm]{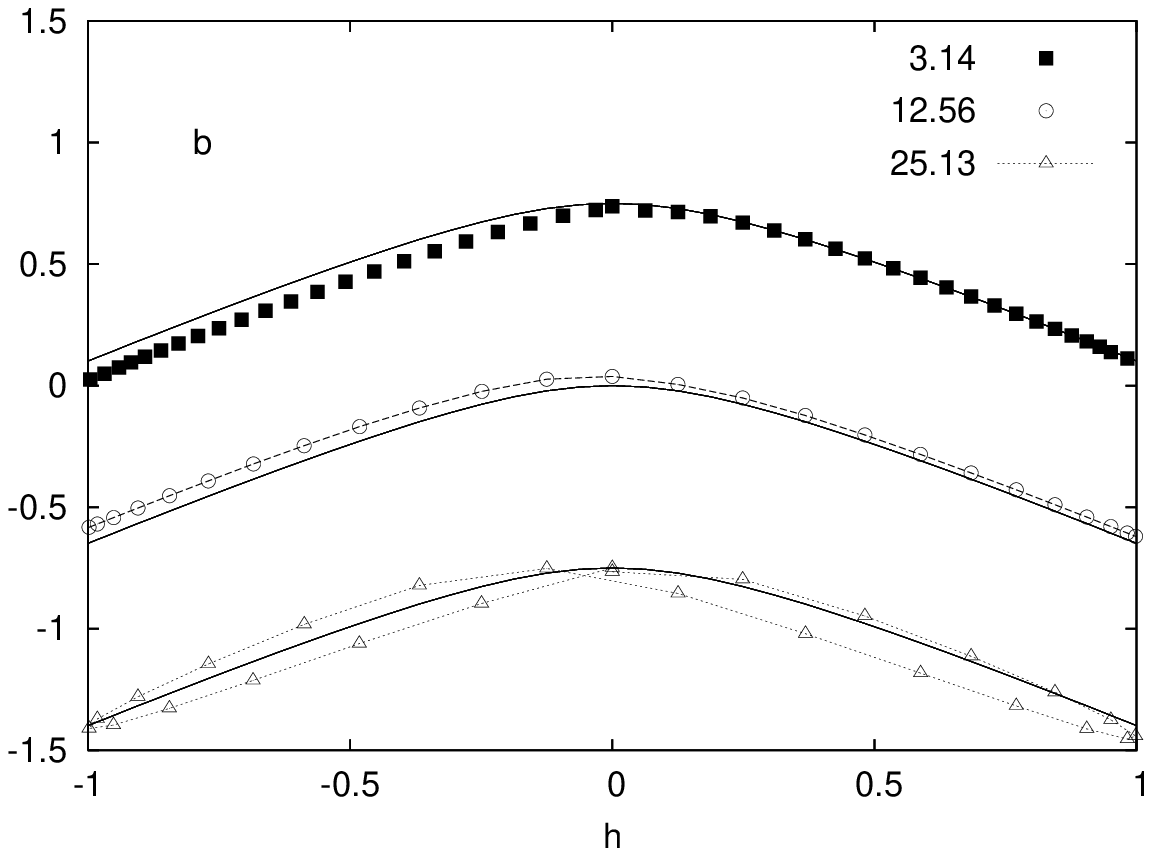}
\caption{Comparison of the equilibrium free energy difference
$\Delta f_t=f_{h(t)}-f_{h_0}$ as a function of $h(t)$ (full line),
with the quantity $F^*_t/N$, as defined by eq. (\ref{fstar}),
obtained with the simulations described in the text. The mean
field interaction parameter $J_0$ is taken to be $J_0=1.1$. In
these simulations  the external magnetic field is varied according
to the oscillatory protocol (\ref{hosc}) with $h_0=1$, and
$\tf=2$. The system size is $N=10$ and  $\Nt=10^4$ samples of the
process are taken. The sets of data are shifted to improve the
clarity of the plot. Panel (b): mean value of of $F^*/N$, obtained
by averaging the contributions to this quantity for a given value
of $h$, as plotted in panel (a). The dotted lines are guides to
the eye.} \label{confrontoj_omega_J1.1}
\end{figure}

We now consider the case $J_0=1.1$ and $\tf=2$. In figure
\ref{confronto_omega_J1.1} the reconstructed free energy landscape
$f^*(m)$ as given by eq.~(\ref{f0star}) is plotted, while in
fig.~\ref{confrontoj_omega_J1.1} the estimated free energy
difference $\Delta F^*_t$, as defined by eq.~(\ref{fstar}), is
plotted. The distance function $d$ for the value of $J_0=1.1$ is
plotted in fig.~\ref{d_omega_tf2}, as a function of the
manipulation protocol frequency $\nu$. Comparison of figures
\ref{d_omega_tf2}, \ref{confronto_omega_J1.1} and
\ref{confrontoj_omega_J1.1} indicates that, for this value of
$J_0$, the optimal frequency is $\nu\simeq 2$, which is smaller
than the value we find for the $J_0=0.5$ case.

\section{Unzipping of a model homopolymer}\label{model_h}
In this section we consider a simple model of homopolymer subject
to external forces. We aim thus to reconstruct the energy
landscape of the polymer, as a function of its internal
coordinate, namely its extension, via non-equilibrium
manipulations.

The model polymer is made up of $N$ identical beads which interact via
a Lennard-Jones potential
\begin{equation}
    U_{LJ}=\sum_{i=1,j<i}^N 4 \epsilon \pq{\p{\frac{\sigma}{r_{ij}}}^{12}
    -\p{\frac{\sigma}{r_{ij}}}^6},
\end{equation}
where $r_{ij}$ is the distance between the $i$-th and the $j$-th
monomers. Successive beads along the polymer chain interact also
via the harmonic potential
\begin{equation}
U_2=\sum_{i=1}^{N-1} k(r_{i,i+1}-\sigma)^2.
\label{u2}
\end{equation}

We use here molecular dynamics simulations with Langevin noise:
the equations of motion of the polymer beads thus read
\begin{equation}
    m \mathbf{\ddot r}_i=\mathbf{F}(\mathbf{r}_i)-\gamma  \mathbf{\dot
    r}_i+\boldsym{\eta}(t),
\end{equation}
where $m$ is the mass of the
bead, $F(\mathbf{ r}_i)$ is the force acting on the i-th bead due
to the interaction with the remaining $N-1$ beads, $\gamma$ is the
friction coefficient and $\boldsym{\eta}(t)$ is the random force
satisfying
\begin{eqnarray}
\average{\boldsym{\eta}(t)}&=&0; \\
\average{\eta_\alpha(t)\eta_\beta(t')}&=&2 k_\mathrm{B}
T\gamma\delta(t-t'), \, \alpha,\beta=x,y,z.
\end{eqnarray}
The values of the model polymer parameters are chosen following
refs.~\cite{VKT,HC}: the Lennard-Jones energy $\epsilon$ and
distance $\sigma$ are taken to be $\epsilon=1$~kcal/mol,
$\sigma=0.5$~nm, respectively, while the monomer mass is taken to
be $m=3\cdot 10^{-25}$~kg. With this choice of the basic
parameters one obtains a characteristic time $\tau\equiv \sqrt{m
\sigma^2/\epsilon}\simeq3.3$~ps. The strength of the harmonic bond
potential (\ref{u2}) is taken to be $k=5000\, \epsilon/\sigma^2$, which corresponds to a more rigid bond than those considered in refs.~\cite{VKT,HC}.
For the friction coefficient we take $\gamma=15 m/\tau$. The
stochastic equations of motion for position and the velocity of
the system's beads are solved using a modified leapfrog algorithm
\cite{Man}, with an integration time step $\delta t=0.005\tau $,
and where the temperature is fixed to $T=300$~K .

\begin{figure}[h]
\center
\psfrag{z}[ct][ct][1.]{$z$}
\psfrag{zeta}[ct][ct][1.]{$\zeta$}
\includegraphics[width=8cm]{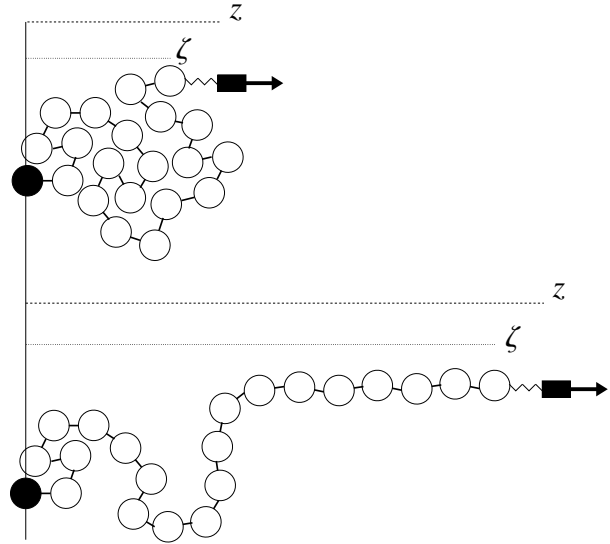}
\caption{Cartoon of the mechanical unfolding of the model
homopolymer. The coordinate $z$ indicates the distance of the
pulling apparatus from the reference plane, while the coordinate
$\zeta$ indicates the distance of the $N$-th monomer from the
reference plane, and represents the system collective coordinate.}
\label{chain}
\end{figure}

In order to mimic the unfolding of the above described system with
an external force exerted by an AFM cantilever, the polymer is
manipulated according to the following procedure: the position of
the first monomer of the chain is kept fixed, mimicking the
trapping in the focus of an optical tweezers of infinite
stiffness; at the starting time the last monomer of the chain is
``attached'' to a pulling apparatus with a spring of elastic
constant $k$ (equal to the ``molecular'' stiffness appearing in
eq.~(\ref{u2})), see figure~\ref{chain}. The external force is
thus applied by moving the apparatus along a fixed direction with
a protocol $z(t)$. Let $\zeta$ denote the distance of the $N$-th
monomer from the plane containing the first monomer and
perpendicular to the applied force direction, the external force
reads thus $F_{ext}=k(z-\zeta)$.

\begin{figure}[h]
\center
\psfrag{t}[ct][ct][1.]{$t/\tau$}
\psfrag{le}[ct][ct][1.]{$\ell\, (\sigma)$}
\includegraphics[width=8cm]{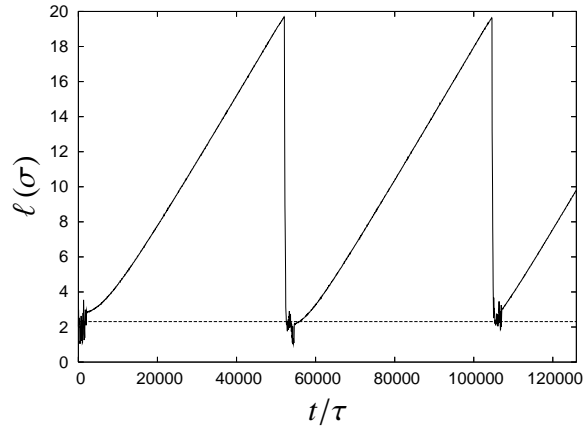}
\caption{Polymer end-to-end length as a function of the time, for
a linear pulling protocol, with velocity $\dot z(t)=5\cdot
10^{-5}\, \sigma/\tau$. The pulling apparatus is detached from the
polymer  after a fixed time $t=50000 \tau$ and then the system
relaxes with no external force applied for a time interval $\Delta
t=2500 \tau$. After that the external force is  applied again.}
\label{sample_etel}
\end{figure}

As expected,  we find that, in the absence of external force, the
model polymer is in a globular state. Let $\ell$ be the end-to-end
distance of the polymer, i.e., the distance between the last and
the first monomer of the chain: $\ell=|\br_N -\br_1|$. We observe
that in absence of external force, this quantity is
$\ell=2.31\pm0.08 \sigma$. In order to define a typical collective
time for the system, we measure the time needed to refold after a
complete unfold, which we take to correspond to an end-to-end
length $\ell=N \sigma$, we define this time $\tF$, which takes the
value $\tF\simeq500 \tau$ for the system size here considered. We
also define the characteristic folding velocity $\vF\equiv N
\sigma/\tF\simeq0.04 \sigma/\tau$. These two quantities define the
intrinsic time and velocity scale of the polymer  dynamics. In
figure~\ref{sample_etel} the end-to-end length is plotted as a
function of the time for a linear pulling protocol, with a
constant velocity $\dot z(t)=5\cdot 10^{-5}\sigma/\tau$ .

We aim now to measure the system intrinsic free energy landscape
as a function of the internal coordinate $\zeta$ using the method
discussed in section \ref{histo_met}. The work done on the polymer
along a single trajectory reads
\begin{equation}
W=\int_0^{\tf} \D t\,  k\pq{z(t)-\zeta(t)}\dot z(t).
\end{equation}
Following equation (\ref{defRs}), the best estimate for
$\FF_0(\zeta)$ is given by  $\FF_0(z)=-k_B T \ln R^*(\zeta)$ where
$R^*(\zeta)$ is given by eq.~(\ref{defRs}), and
$U_{\mu(t)}(\zeta)=\frac k 2\pq{z(t)-\zeta}^2$. As in the previous
section we consider here both a linear protocol and an oscillatory
protocol.

\subsection{Linear protocol}\label{poly_lin:subsec}
In this section we consider the linear pulling protocol
\begin{equation}
z(t)=z_0+v t,
\label{pol_lin}
\end{equation}
where the constant $z_0$ is chosen to be slightly greater than the z-position
of the N-th monomer at the beginning of each trajectory: $z_0=\zeta(t=0) +\sigma/100$.
Here we consider  three values of the pulling velocity, $v=5\times 10^{-4},\, 5\times 10^{-3},\,5\times 10^{-2}\, \sigma/\tau$. For each velocity, the duration time of the manipulation $\tf$ is chosen in such a way that the stroke of the pulling apparatus is $\Delta z=25 \sigma$, and the polymer is fully unfolded. This corresponds to a time interval of $\tf=500000,50000, 5000 \tau$ for the the three  protocols, respectively.
For the slowest velocity we take 50 repetitions of the pulling process, for
the intermediate velocity  we take 500 repetitions,
while for the fastest velocity we take 1000 repetitions of the pulling process.
After each pulling process, the polymer  evolves at zero force for a time interval of $5\tF=2500 \tau$, mimicking the detachment of the pulling apparatus and the refolding of the polymer.

\begin{figure}[h]
\center
\psfrag{effe}[ct][ct][1.]{ $\FF^*_0\, (k_B T)$}
\psfrag{zet}[ct][ct][1.]{ $\zeta$}
\psfrag{5e-4}[cl][cl][.8]{$5\times 10^{-4}\, \sigma /\tau$}
\psfrag{5e-3}[cl][cl][.8]{$5\times 10^{-3}\, \sigma /\tau$}
\psfrag{5e-2}[cl][cl][.8]{$5\times 10^{-2}\, \sigma /\tau$}
\includegraphics[width=8cm]{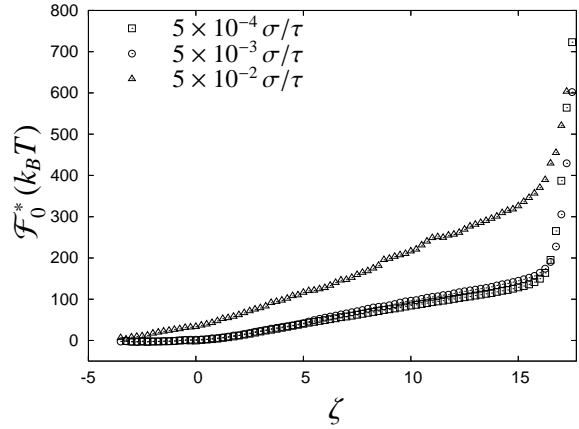}
\caption{Reconstructed free energy landscape $\FF^*_0$ as a function of the polymer internal coordinate $\zeta$, obtained with the linear protocol (\ref{pol_lin}), for the three pulling velocities here considered. The line, in the case of the faster velocity, is a guide to the eye.}
\label{complin}
\end{figure}

\begin{figure}[h]
\center
\psfrag{F}[ct][ct][1.]{ $\Delta F\, (k_B T)$}
\psfrag{x}[ct][ct][1.]{$z$}
\includegraphics[width=8cm]{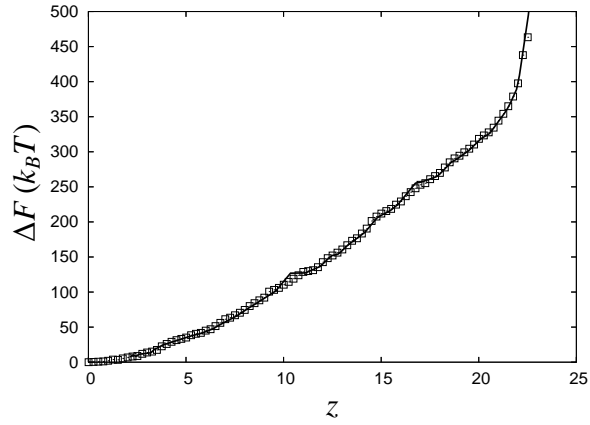}
\caption{Free energy difference $\Delta F$ as a function of the pulling apparatus coordinate $z$, for the linear protocol (\ref{pol_lin}), with the largest pulling velocity here used  $\dot z=5\times 10^{-2} \sigma/\tau$.
Squares: free energy difference $\Delta F$ obtained using the JE (\ref{fstar}). Full line: free energy difference $\Delta F$ as given by eq.~(\ref{comp_eq}).}
\label{conf_poly}
\end{figure}

In fig.~\ref{complin} we compare the free energy landscape  $\FF^*_0(\zeta)$ obtained from the three pulling velocities, by using the histogram method discussed in section \ref{histo_met}. Differently form the Ising model, in this case, we do not know the
expected free energy function. However, in order to perform a consistency check 
one can note that the  free energy difference $\Delta F_{z(t)}=F(z(t))-F(z(0))$, which is function of the pulling apparatus coordinate $z(t)$, and the free energy landscape  $\FF_0(\zeta)$ are related via
\begin{equation}
 \Delta F_{z(t)}=-k_B T \ln \pq{\int \D \zeta\, e^{-\beta\pq{\FF_0(\zeta)+\frac {k}{2} \p{z(t)-\zeta}^2}}} +\mathrm{const}.
\label{comp_eq}
\end{equation}
Our best estimate for $\Delta F_{z(t)}$ is obtained by averaging $\exp\p{-\beta W_t}$ over the repetitions of the pulling process, as given by eq.(\ref{fstar}).
In the limit of $\dot z \rightarrow 0$, we expect this estimate to be exact.
In figure \ref{conf_poly}, we compare the free energy difference obtained with direct measuring, eq.~(\ref{fstar}), and that obtained using eq.~(\ref{comp_eq}), for the fastest pulling velocity here used.   Inspection of this figure suggests that the agreement between the two estimates of $\Delta F_{z(t)}$ is rather good for this value of the velocity. The agreement is also good for the two other  values of the pulling velocity (data not shown).

\subsection{Oscillatory protocol}
\begin{figure}[h]
\center
\psfrag{te-1}[ct][ct][1.]{$t/\tau\, 10^{-1}$}
\psfrag{le}[ct][ct][1.]{$\ell\, (\sigma)$}
\psfrag{a}[ct][ct][1.]{$(\mathrm{a})$}
\psfrag{b}[ct][ct][1.]{$(\mathrm{b})$}
\includegraphics[width=8cm]{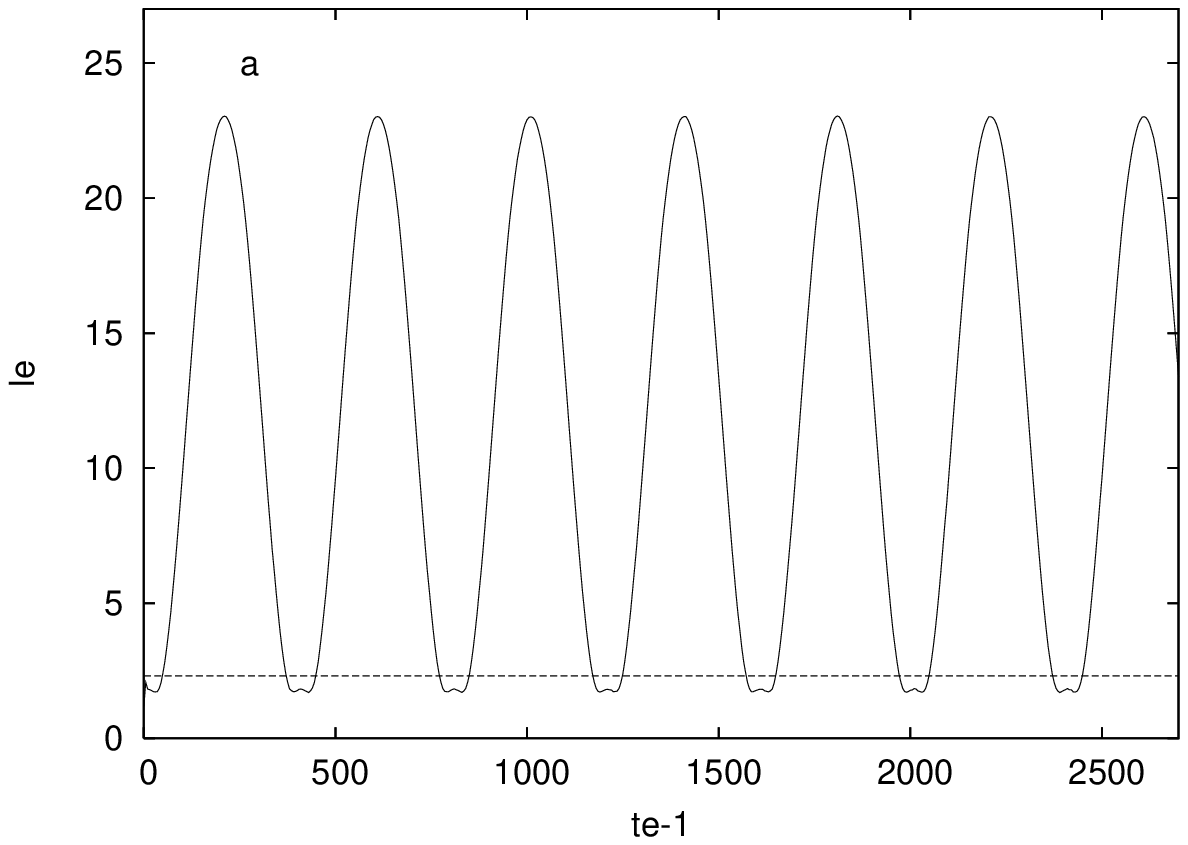}
\includegraphics[width=8cm]{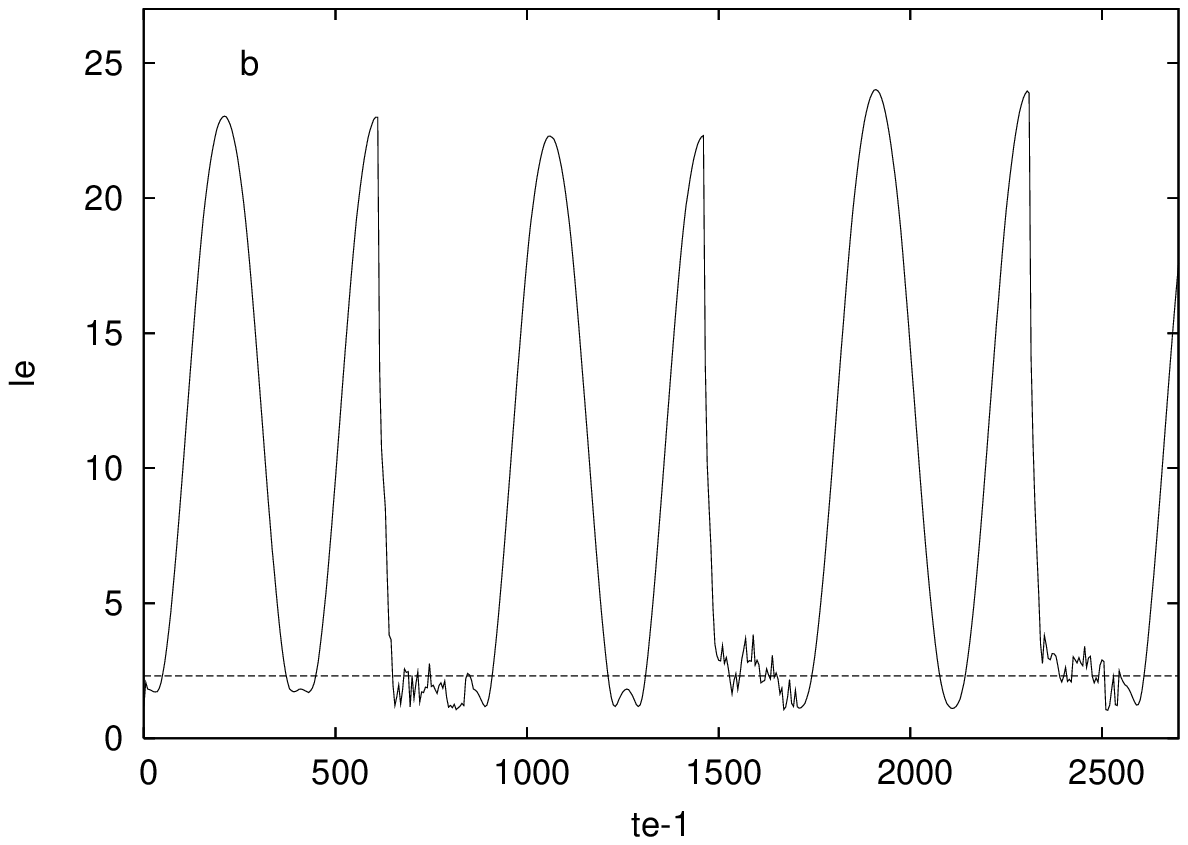}
\caption{Polymer end-to-end length as a function of the time, for
the periodic manipulation protocol (\ref{osc_pol}), with frequency
$\nu=5 \times 10^{-4} \tau^{-1}$. The dotted line represents the
mean value of the polymer end-to-end length in absence of external
force. Panel (a): the pulling apparatus stays in contact with the
polymer during the whole manipulation process. Panel (b): the
pulling apparatus  is detached from the polymer after a fixed time
$1.5/\nu \,  \tau$ and then the system relaxes with no external
force applied for a time interval $\Delta t=2500 \tau$. After that
the external force is  applied again.} \label{sample_osc_etel}
\end{figure}

\begin{figure}[h]
\center
\psfrag{z}[ct][ct][1.]{$z$}
\psfrag{x}[ct][ct][1.]{$x$}
\psfrag{zeta}[ct][ct][1.]{$\zeta$}
\includegraphics[width=8cm]{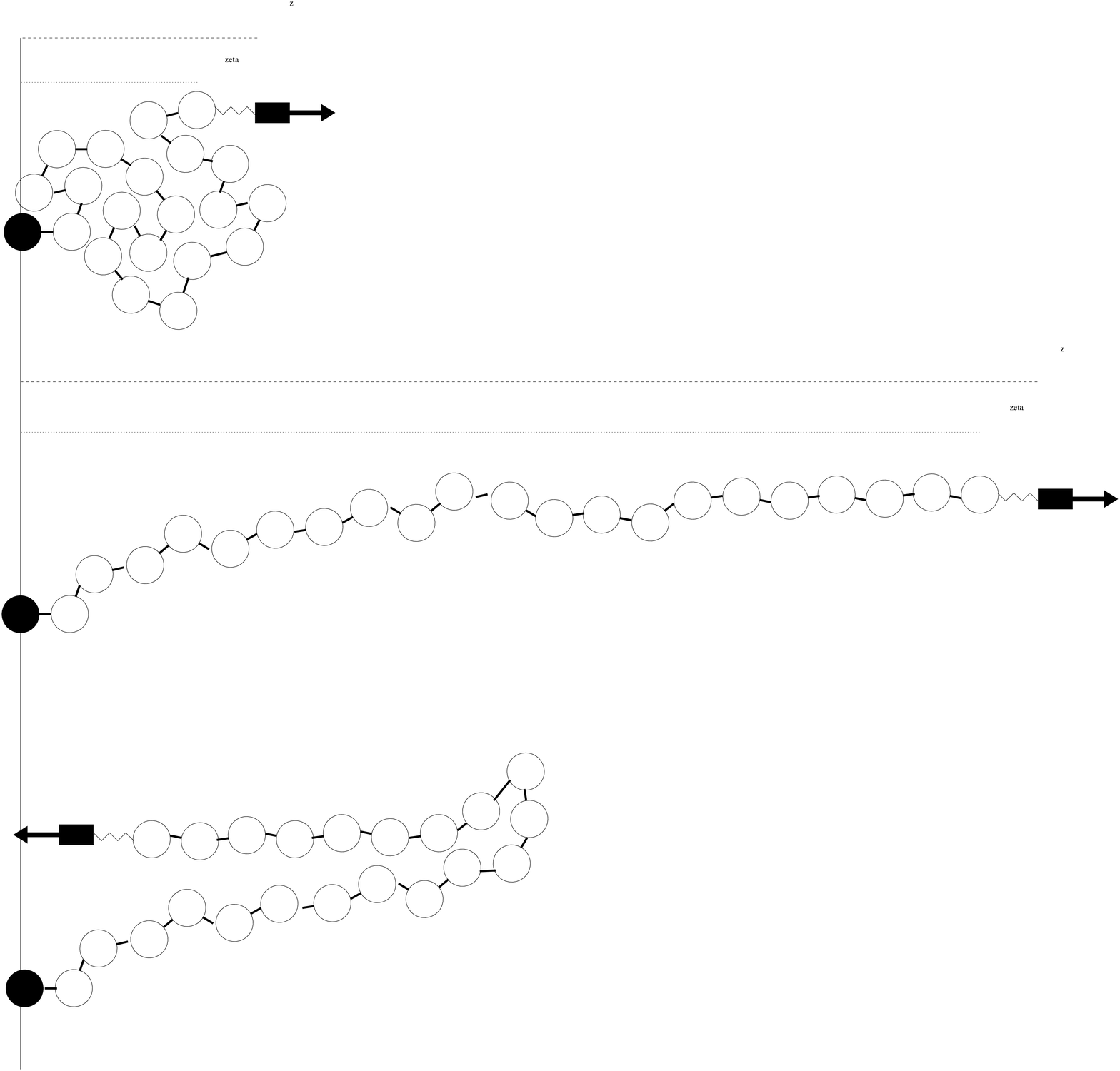}
\includegraphics[width=8cm]{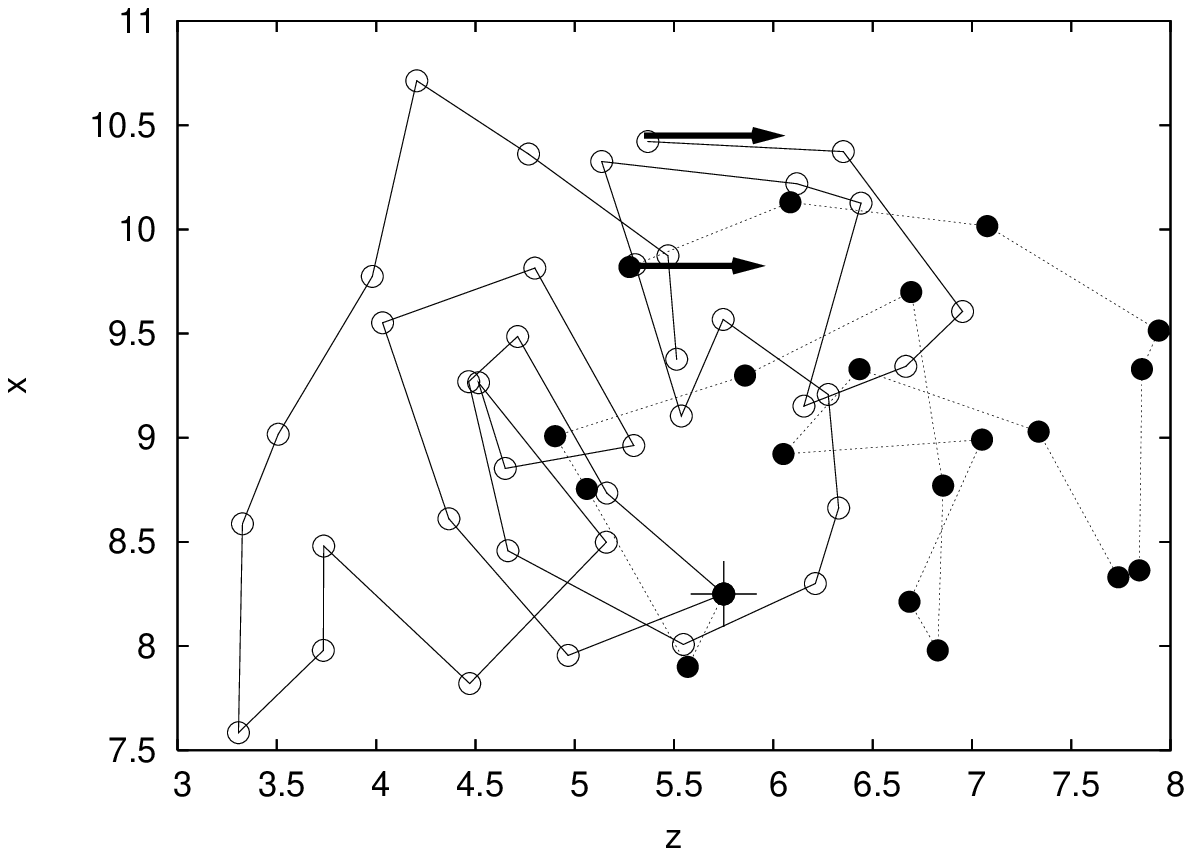}
\caption{Upper panel:Cartoon of the polymer manipulation. The polymer is manipulated with the oscillatory protocol (\ref{osc_pol}). If the polymer is manipulated too fast, it cannot attain the globular native state after one manipulation cycle.
Lower panel: snapshots of the polymer configuration, at two different times of a simulation run with $\nu=5 \times 10^{-4}\, \tau^{-1}$, projected onto the $z-x$ plane. Empty circles-full line: Configuration at $t=0$, i.e. the system is
at thermodynamical equilibrium with no external force applied. Full circles- dotted line: configuration of the system after one manipulation period. The cross indicates the fixed monomer, the arrows indicate the monomers to which the force is applied. }
\label{poly1}
\end{figure}

Here, the polymer is manipulated by varying the pulling apparatus position according to
the protocol
\begin{equation}
z(t)=\frac{\zm}{2} \pq{1-\cos(2 \pi \nu t)}+z_0,
\label{osc_pol}
\end{equation}
where the constant $z_0$ is chosen to be slightly greater than the z-position
of the N-th monomer at the beginning of each trajectory: $z_0=\zeta(t=0) +\sigma/100$.
The value of $\zm$ is taken to be $\zm=(N+4)\sigma$, in such a way that the polymer is fully unfolded near the maximum of the function (\ref{osc_pol}).
Three values of the frequency $\nu=5\times 10^{-6}, 5\times 10^{-5},\,  5\times 10^{-4}\,  \tau^{-1}$ are considered here. This values have to be compared with the system characteristic frequency $\nu_{\mathrm F}$ as estimated in section \ref{model_h}, $\nu_{\mathrm F}=1/\tF\simeq 2 \times 10^{-3} \tau^{-1}$.
Let us define the effective velocity
\begin{equation}
\zeff\equiv \p{\nu \int _0^{1/\nu} d t\,  \dot  z^2(t)}^{\frac 1 2 }=\frac{\zm \pi \nu}{\sqrt 2},
\end{equation}
the three frequencies here considered correspond to
 the values of the effective velocity
$\zeff\simeq2.66 \times 10^{-4},\, 2.66 \times 10^{-3},\,  2.66 \times 10^{-2}\, \sigma/\tau$ respectively. These velocities have to be compared with the characteristic folding velocity of the polymer as estimated in section \ref{model_h}, $\vF=0.04\, \sigma/\tau$.
We adopt two different approaches to manipulate the polymer: in the first case the pulling apparatus is always attached to the polymer during the whole manipulation time $\tf$, fig.~\ref{sample_osc_etel}(a), while in the second case  the pulling apparatus is detached after one and half period $1/\nu$ of the protocol, the system equilibrates at zero force for  a time interval of $5 \tF$, and then the force is applied again, see figure \ref{sample_osc_etel}(b).
In the case where the pulling apparatus is attached during the whole manipulation process, we take a total manipulation time $\tf=10^7\, \tau$. In the case where the pulling apparatus is detached after one and half period, we consider 50 trajectories for $\nu=5 \times 10^{-6}$, 500 trajectories for $\nu=5 \times 10^{-5}$, and 1000 trajectories for $\nu=5 \times 10^{-4}$.
\begin{figure}[h]
\center
\psfrag{effe}[ct][ct][1.]{ $\FF^*_0\, (k_B T)$}
\psfrag{zet}[ct][ct][1.]{ $\zeta$}
\psfrag{w5e-6}[cl][cl][.8]{$\nu=5\times 10^{-6}$}
\psfrag{w5e-5}[cl][cl][.8]{$\nu=5\times 10^{-5}$}
\psfrag{w5e-4}[cl][cl][.8]{$\nu=5\times 10^{-4}$}
\includegraphics[width=8cm]{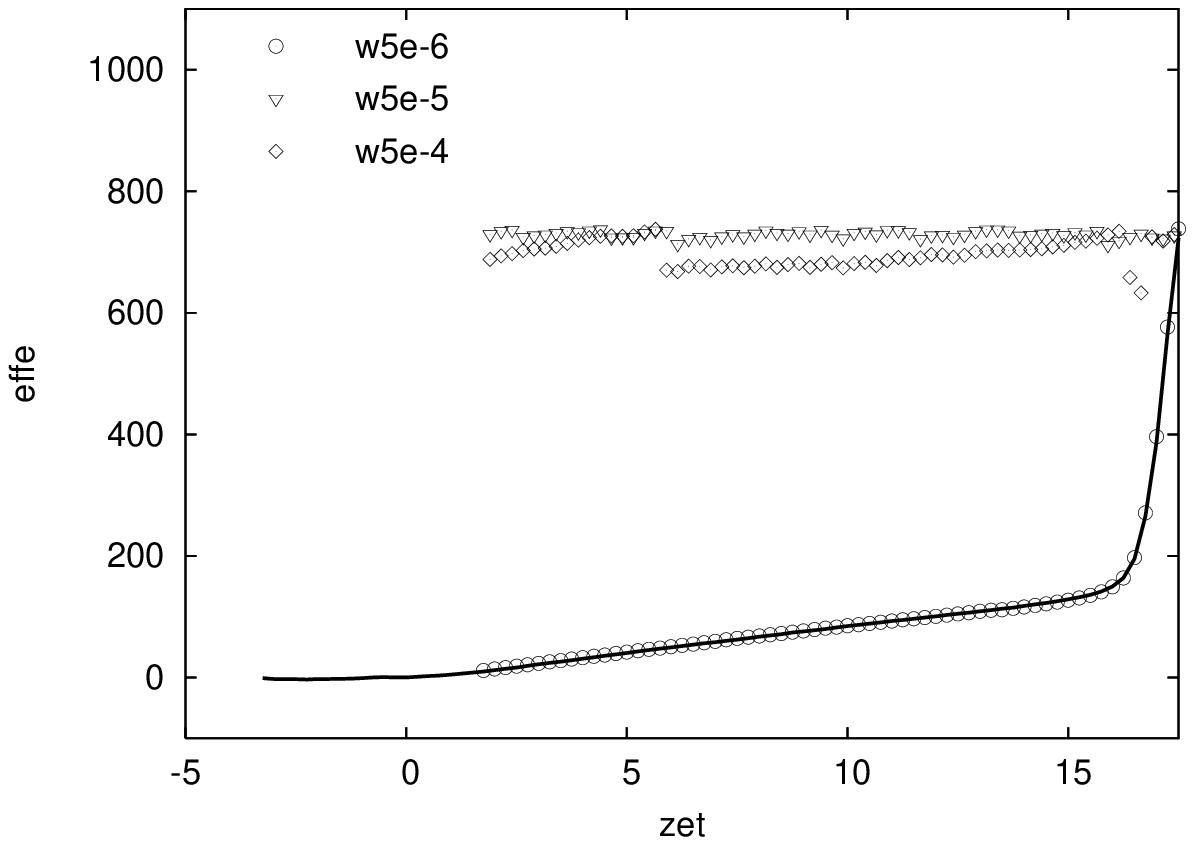}
\includegraphics[width=8cm]{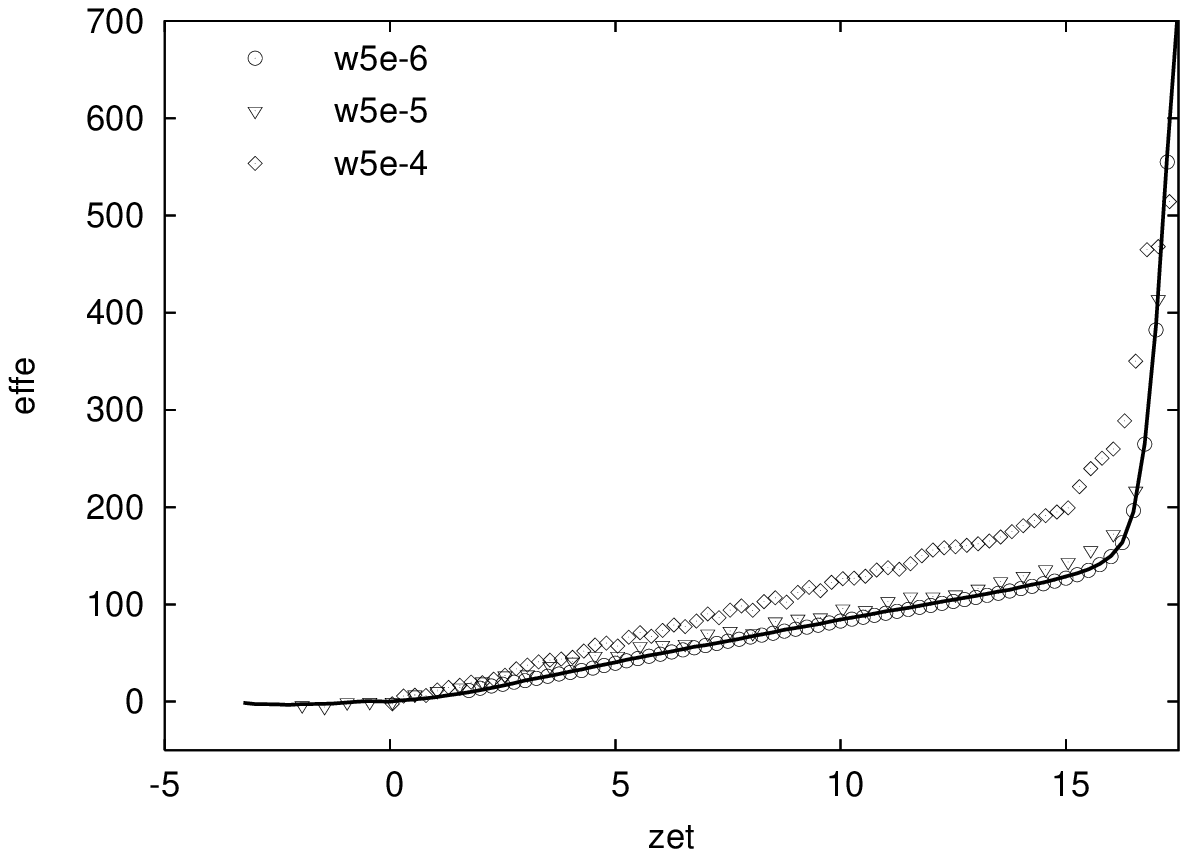}
\caption{Reconstructed free energy landscape $\FF^*_0$ as a function of the polymer internal coordinate $\zeta$, obtained with the oscillatory protocol (\ref{osc_pol}) for the three frequency $\nu$ here considered. The full line is the ``reference'' free energy landscape $\FF_0$ obtained  with the linear protocol (\ref{pol_lin}) with the smallest velocity $v=5\times 10^{-4}\, \sigma/\tau$. Upper panel: the free energy landscape $\FF_0$ is reconstructed using the ``always-attached'' manipulation protocol, fig.~\ref{sample_osc_etel}(a). Lower panel:  the free energy landscape $\FF_0$ is reconstructed by periodically detaching the pulling apparatus, fig.~\ref{sample_osc_etel}(b).}
\label{rec_osc}
\end{figure}
The results for the reconstructed free energy landscape with these two manipulation strategies are plotted in fig.~\ref{rec_osc}. Inspection of figure \ref{rec_osc}(a) clearly puts in evidence that the ``always attached'' protocol (fig.~\ref{sample_osc_etel}(a)) gives a good estimates for the free energy landscape only for the smallest frequency here considered $\nu=5\times 10^{-6}\, \tau^{-1}$, which corresponds to an effective velocity $\zeff=2.66 \times 10^{-4}\, \sigma /\tau$; while for the two other frequencies the reconstructed free energy landscape is completely wrong.
This can be easily understood by looking at fig.~\ref{poly1}: after a complete unfolding, if the molecule is pulled leftwards too fast, it cannot achieve the native globular state, and so the internal coordinate $\zeta$, will be no longer a ``good'' collective coordinate to describe the system state.
Note that the periodic protocol proves unsuccessful to recover the energy landscape for frequency well below the system characteristic frequency $\nu_{\mathrm F}$.

In the case of the second manipulation strategy, the reconstructed energy landscape, fig. \ref{rec_osc}(b), agrees with that obtained with the linear protocol
for the two smallest frequency here considered. On the contrary, the reconstructed energy landscape obtained with the largest frequency $\nu=5 \times 10^{-4}$ is clearly inaccurate.

It is  worth to note that for our purpose, i.e. the reconstruction of 
the free energy landscape, the ``pulsed" protocol, represented in figure \ref{sample_osc_etel}(b) is to all extent equivalent to the linear protocol 
(\ref{pol_lin}).

\section{Discussion}\label{discuss}
In the present work we have combined the extended form
of the JE, eq.~(\ref{sample}), and  the histogram method to reconstruct the free energy landscape of two simple systems driven out of equilibrium by manipulation of an external parameter.

In the case of the Ising model in  mean field approximation, the external magnetic field is manipulated both with a linear and with a periodic protocol.
In both cases, for a sufficiently gentle protocol, the system free energy landscape is successfully evaluated. It is worth to note that, for the periodic protocol, the optimal frequency for the reconstruction of the landscape is somewhat larger than the smallest frequency here considered. This indicates the existence 
of a typical system frequency, which optimizes the estimate given by the histogram method, as already found in \cite{Seif1}. However, we point out that this 
typical frequency is of the order of the frequency governing the system dynamics, which has been taken equal to one in the present work. This means that the 
manipulation has to be performed on time scales similar to the system characteristic time scale, in order for  the free energy evaluation method  here discussed to be successful. Faster manipulations give unreliable estimates of the free energy function.
In the case of the linear protocol we also consider the effect of the system size on the effectiveness of the histogram method. As discussed in refs.~\cite{noi2,noi3}, changing the system size $N$ corresponds to change the system energy scale. There, we showed that one can obtain a good estimate of the free energy difference, via the JE, only for small system sizes (small energy scales). Similarly, the results of the present works indicate that the histogram method
is effective for small system sizes. This conclusion widens the results of 
refs.~\cite{HumSza,Seif1} on the histogram method,  since in those references the effect of the energy scale was not considered.

The second system here considered is a simple model of homopolymer, which is 
unzipped by applying an external force to one of its free ends.
Also in this case the external force is varied both with a linear and with a periodic protocol.
The results of this simulated experiment have to be considered more carefully, with respect to the case of the Ising model, 
since we do not know the exact shape of the polymer free energy landscape.
We take as our best estimate of this landscape the one provided by the linear protocol 
with the smallest velocity.
We find that the periodic force gives  unreliable estimates of the free energy as a function of 
the polymer elongation, even for frequencies much smaller than the system characteristic frequency.
This is at variance with the conclusions of ref.~\cite{Seif1}, where the periodic loading was found to be the optimal one for the evaluation of the free energy landscape of a model polymer.
The reason for this discrepancy resides in the fact that our model polymer
takes also into account the three-dimensional structure of the system, and 
when a periodic force is applied, the elongation coordinate is no longer
a ``good'' collective coordinate, and fails to catch the connection 
between the system macroscopic state and its microscopic state, as depicted in fig.~\ref{poly1}.
In fact, the system has no time to recover its initial globular state, and keeps memory of previous trajectories at each manipulation cycle.

Our results suggest thus that in the realization of a real experimental set-up, if one wants to exploit the histogram method to evaluate the free energy landscape of a polymer, 
some care has to be taken with the choice of the manipulation protocol.
The linear protocol, or the ``pulsed'' sinusoidal protocol, appear to be the best choices to this purpose. This is closely related to the original proposition of the JE, which states that the equality holds {\it if} the system is in thermodynamical equilibrium at the beginning of the manipulation.

Finally, we have found that the polymer elongation is not a good
state variable to describe the system when the manipulation occurs
on too short times, since the system is not able to reach a
quasi-equilibrium state defined by its instantaneous value. In
this case, the concept of a free-energy landscape depending on
this coordinate is ill-defined.

\ack
This research
was partially supported by MIUR-PRIN 2004.

\end{document}